\documentclass{epl}
\usepackage[T1]{fontenc}
\title{Antiferromagnetic interlayer coupling in ferromagnetic semiconductor \chem{EuS/PbS(001)} superlattices}
\shorttitle{Antiferromagnetic interlayer coupling etc.}
\author{H. K\k{e}pa\inst{1,6}\thanks{E-mail: Henryk.Kepa@fuw.edu.pl} 
\and J. Kutner-Pielaszek\inst{1} \and 
J. Blinowski\inst{1} \and A. Twardowski\inst{1} \and C.F. Majkrzak\inst{2}
\and  T.~Story\inst{3} \and P. Kacman\inst{3} \and R.R.~Ga{\l}\k{a}zka\inst{3} 
\and K. Ha\inst{4} \and H.J.M. Swagten\inst{4} \and W.J.M. de Jonge\inst{4} 
\and
A.Yu. Sipatov\inst{5} \and V. Volobuev\inst{5} \and T.M. Giebultowicz\inst{6}}
\shortauthor{H. K\k{e}pa \etal}
\institute{
  \inst{1} Department of Physics, Warsaw University -- ul.~Ho\.za~69, 00-681
Warszawa, Poland\\
  \inst{2} National Institute of Standards and Technology -- Gaithersburg,
 MD 20899, USA\\
  \inst{3} Institute of Physics, Polish Academy of Sciences -- Al.~Lotnik\'ow
32/46, 02-668 Warszawa, Poland\\
  \inst{4} Department of Applied Physics and COBRA -- Eindhoven University of
Technology, 5600 MB Eindhoven, The Netherlands\\
  \inst{5} Kharkov State Polytechnical University -- 21 Frunze St., 310002
Kharkov, Ukraine\\
  \inst{6} Physics Department, Oregon State University -- Corvallis, OR 97331,
USA
}

\begin{document}

\maketitle

\begin{abstract}
Antiferromagnetic coupling between ferromagnetic layers has been observed
for the first time in an all-semiconductor superlattice structure
\chem{EuS/PbS(001)}, by neutron scattering and magnetization measurements.
Spin-dependent superlattice band structure effects are invoked to explain
the possible origin and the strength of the observed coupling. 
\end{abstract}

Since the discovery of antiferromagnetic (AFM) coupling
between  metallic
ferromagnetic (FM) layers separated by a nonmagnetic metal
in late 1980s, the
subject has been an active research area in fundamental magnetism. The AFM
coupling plays a key role in many technological applications such as
magnetoresistive sensors and magnetooptical devices \cite{prinz}.
In metallic systems, the interlayer
coupling results from the quantum interference of conduction
electron waves in the spin
dependent potential of magnetic multilayer or, equivalently, via
the spin polarization of conducting carriers (an analog of the
well-known  RKKY interaction \cite{{bruno},{jones}}).
An important question arises as to how interlayer coupling can exist in
non-metallic systems.
Recently, efforts have been  made to study the interlayer coupling in hybrid
systems composed of metallic ferromagnets separated by semiconductor layers.
In particular, \chem{Fe/Si} trilayers and multilayers had attracted much 
attention,
but the coupling turned out to proceed via metallic iron silicide, and
not via the silicon, due to intermixing between iron and silicon occurring
even at room temperature \cite{strijkers}.
This is one illustration of the severe limitations brought
about by the technological incompatibility of FM metals and
semiconductors.
There is no such limitation in the case of the magnetic/nonmagnetic
multilayers composed of properly chosen all-semiconductor materials.
By means of neutron diffraction, an interlayer exchange interaction was
observed
in  {\em semiconductor antiferromagnetic/nonmagnetic}
\chem{EuTe/PbTe}, \chem{MnTe/ZnTe} and \chem{MnTe/CdTe} superlattices (SLs)
\cite{{kepa},{rhyne},{nunez}}.
The {\em ferromagnetic/nonmagnetic semiconductor} \chem{GaMnAs/AlGaAs} system
\cite{chiba} was also studied where a
{\em ferromagnetic}  interlayer exchange was detected in magnetic hysteresis
measurements. Here the coupling is believed to be
mediated, as in metals, by conducting carriers (holes) in the  nonmagnetic
layer, though the very origin of ferromagnetism in \chem{GaMnAs}
\cite{{ohno},{dietl}} is an on-going debate.
Recently developed  models of interlayer coupling specific to the
FM semiconductor multilayers, considered in
\cite{{shevchenko},{dugaev},{blin}}, lead also to the FM correlations.

In this Letter, we present the first conclusive evidence for the existence of
AFM  interlayer coupling in all-semiconductor
ferromagnet/diamagnet \chem{EuS/PbS(001)} multilayers, based on 
neutron diffraction,
neutron reflectivity, and magnetization measurements.
AFM correlations in \chem{EuS/PbS} multilayers are
seen for \chem{PbS} thickness up to 90~\AA. Since AFM interlayer coupling is a
crucial element for the
operation of many important magnetic thin film structures, such as
spin valves, the present finding is of potential technological as well
as fundamental scientific interest.

\begin{figure}
 \twofigures[scale=0.65]{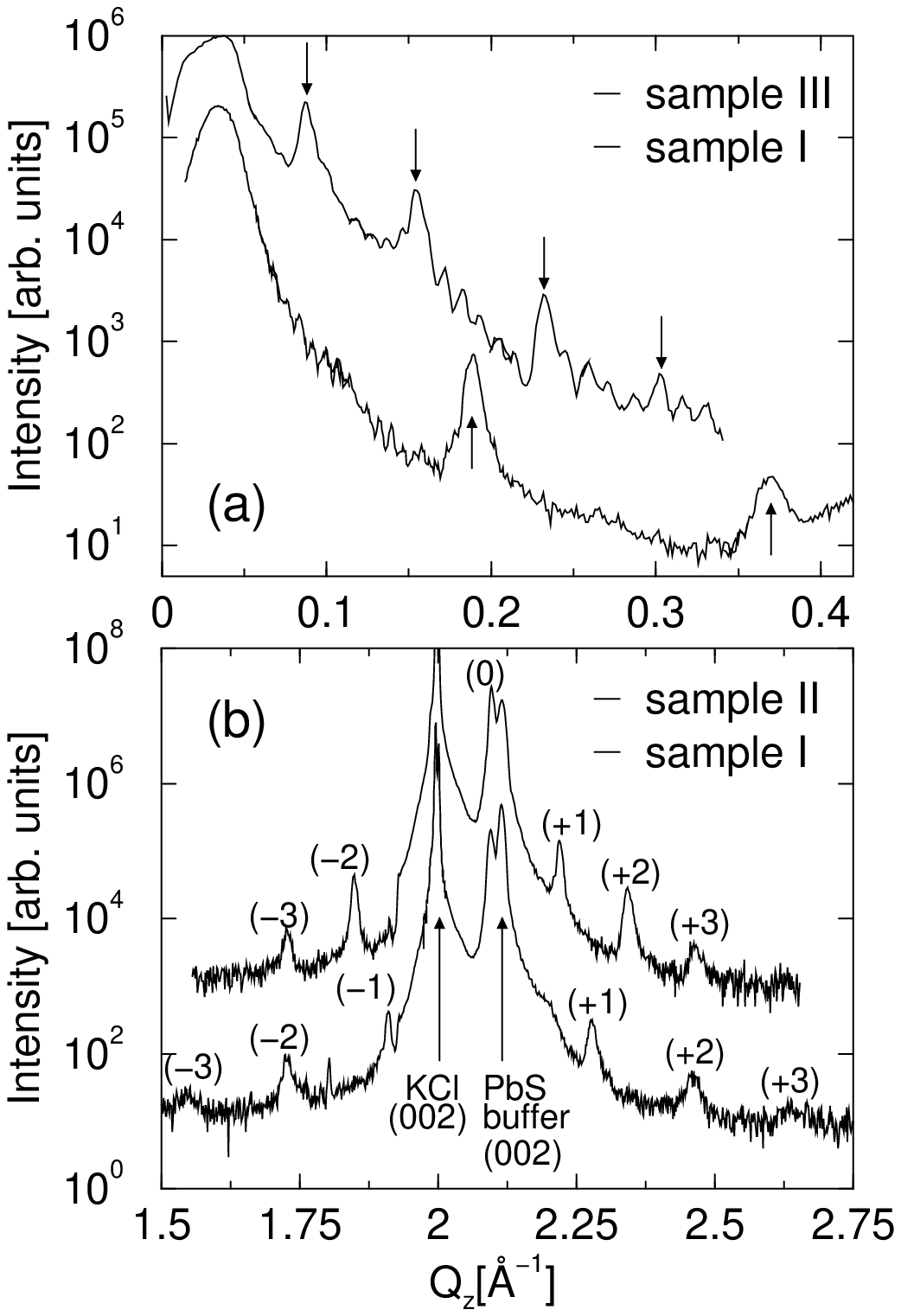}{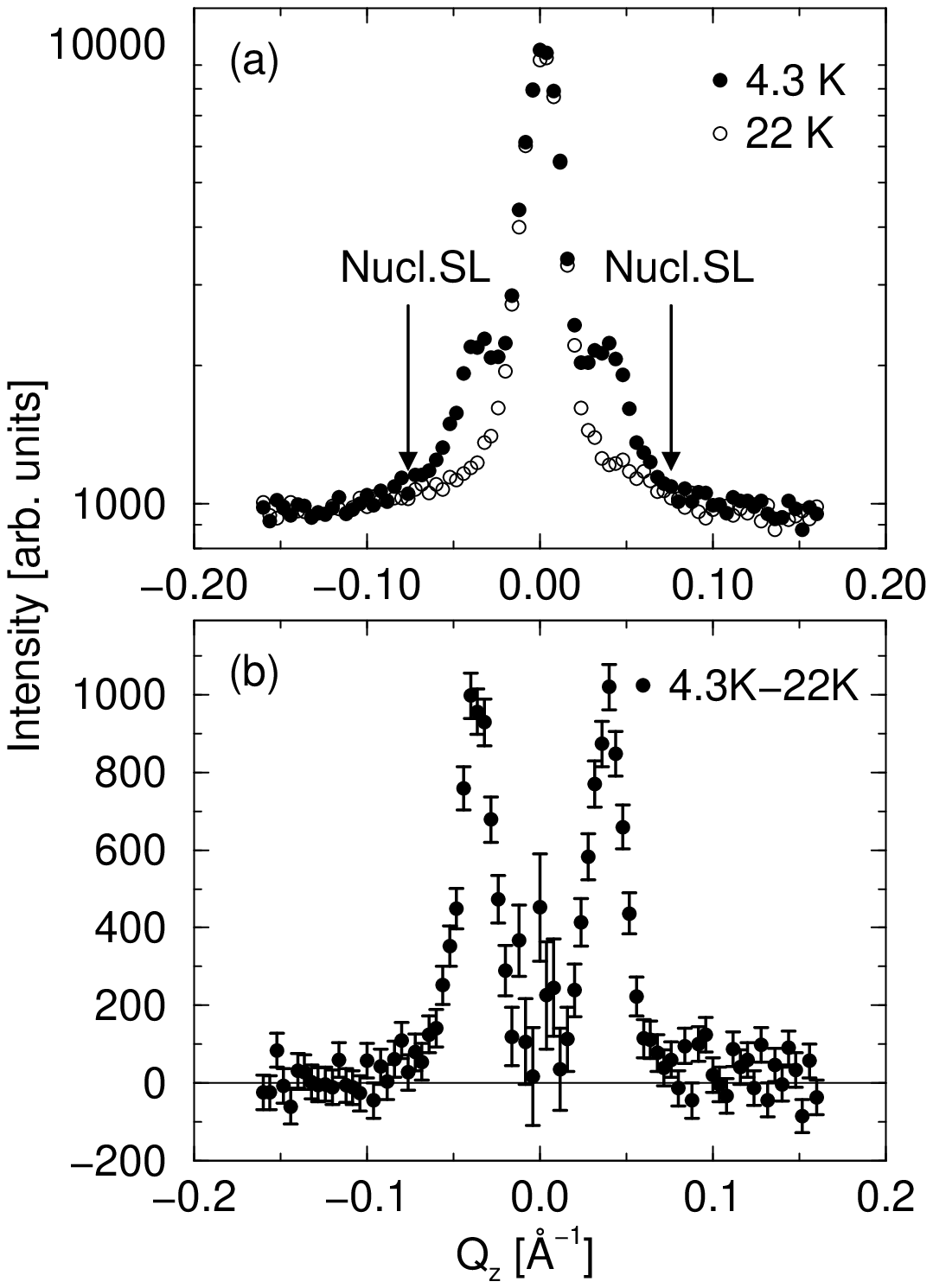}
 \caption{$(a)$ X-ray reflectivity profiles taken for the samples III and I.
  Total reflection regions and a number of superlattice Bragg peaks 
  (marked by arrows) are
  clearly visible. $(b)$ Wide angle diffraction spectra taken about (002)
  reciprocal lattice point for the samples II and I. Apart from 
  strong $(002)$ Bragg reflections from
  the \chem{KCl} substrate and \chem{PbS} buffer layer, large 
  number of smaller SL
  sattelites is also present proving the high structural quality of the
  investigated samples.}
 \label{xray}
 \caption{$(a)$ Neutron diffraction scan with $Q_z$ parallel to the SL
  $[001]$ growth
  direction about  the reciprocal lattice point $(0,2.14,Q_z)$ above and below
  $T_{\rm C}$ for sample III. The arrows in $(a)$ shows the expected positions
  of the first order nuclear SL satellites.
  $(b)$ The purely magnetic contribution is found
  by taking the difference between the two spectra in $(a)$. }
 \label{bt9}
 \end{figure}

Although many samples with different \chem{EuS} and \chem{PbS} thicknesses
(30\AA -- 80\AA \hspace{1mm} and 4\AA -- 90\AA, respectively) have been
examined for this study, we will focus on the results of three
representative samples, which we shall label as
samples I, II, and III.  The compositions of the samples are
[30\AA~\chem{EuS}/4.5\AA~\chem{PbS}]$\times$20,
[35\AA~\chem{EuS}/10\AA~\chem{PbS}]$\times$10, and
[60\AA~\chem{EuS}/23\AA~\chem{PbS}]$\times$15 respectively.
\chem{EuS} and \chem{PbS} are semiconducting materials which
crystallize in the  rocksalt structure with a lattice mismatch of about
0.5$\%$. EuS is a well-known nonmetallic (semi-insulating) Heisenberg
ferromagnet.  \chem{PbS} is a diamagnet with a narrow band gap; its carrier
concentration is typically of the order of 10$^{17}$ to 10$^{18}$ cm$^{-3}$.
The multilayers were grown epitaxially on monocrystalline \chem{KCl(001)} 
substrate
and several hundred angstroms thick \chem{PbS} buffer
using electron beam for \chem{EuS} evaporation and standard resistive 
heating for \chem{PbS}
evaporation.  The quality of the SLs  was examined by X-ray and
neutron diffraction. Figure \ref{xray}$(a)$ shows X-ray reflectivity
profiles for the samples III and I. Four order of SL Bragg peaks are visible
for the sample III, and two order are shown for the sample I, the latter one
having much smaller SL periodicity. In Figure \ref{xray}$(b)$ high angle 
diffraction spectra for the sample II and again the sample I are presented.
Large number of SL sattelites visible in both cases clearly indicates the
high degree of the superlattice structural perfection.
Detailed studies of the growth and magnetic
properties of \chem{EuS/PbS} multilayers with thick \chem{PbS}
spacers  (magnetically decoupled case) has been reported
elsewhere \cite{stachow}.

Neutron diffraction and reflectivity measurements were carried
out at the NIST Center for Neutron Research (NCNR) in Gaithersburg (USA).
A triple-axis spectrometer (neutron wavelength $\lambda=2.35$ \AA)
was set to the elastic mode of operation for the
diffractometry experiments.
Fig.~\ref{bt9}$(a)$ shows a diffraction scan with $Q_{z}$ parallel to the SL
[001] growth direction for sample III in zero magnetic field.
Measurements above and below the Curie temperature $T_{\rm C}$ = 18.5$\pm$
0.5~K (22~K and 4.3~K, respectively) were taken in order
to separate the magnetic contribution from the structural part.
The central peak ($Q_{z}=0$) corresponds to the common (\chem{EuS} and 
\chem{PbS}) nuclear 
in-plane $(020)$ Bragg reflection 
of the whole epitaxial structure grown on the \chem{KCl} substrate.
The smaller satellites visible only in the spectrum taken at 4.3 K are
a signature of the AFM ordering between adjacent EuS layers.  These
peaks occur at $Q_{z}$ = $\pm$ 0.038 \AA$^{-1}$, which  is
about $\pm 2\pi/2d$ where $d$ is the bilayer repeat thickness of the
SL ($d$ = 83~\AA \hspace{1mm} for this sample).
The expected positions of the first-order structural (nuclear) SL satellites
are at $Q_{z}=\pm$ 0.076 \AA$^{-1}$. They are not visible in the scans
due to the combined effect of the superlattice chemical structure 
factor, which has minima in the vicinity of $Q_{z}\approx\pm0.07$ 
\AA$^{-1}$, and considerably smaller nuclear scattering length density (SLD) 
contrast between \chem{EuS} and \chem{PbS} than the magnetic SLD contrast. 
Similar arguments 
can be applied for the reflectometry measurements described below.
The pure magnetic contribution to the scattering as shown in
Fig.~\ref{bt9}$(b)$
is obtained by taking the difference of the two spectra in Fig.~\ref{bt9}$(a)$.
The absence of any magnetic scattering at the central peak position
and the presence of the satellites exactly halfway between the
central peak and the expected first-order nuclear  peak are clear evidence of
AFM interlayer correlations in this \chem{EuS/PbS} system.

\begin{figure}
 \onefigure[scale=0.7]{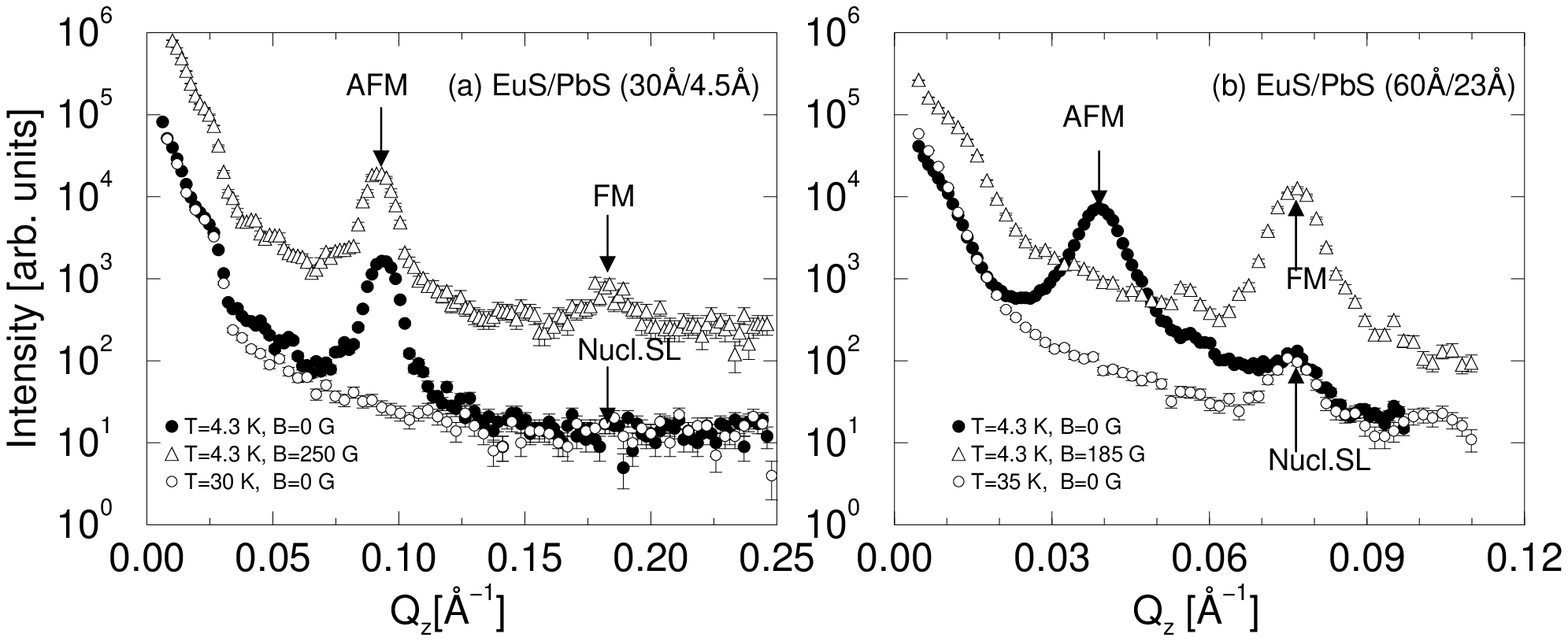}
 \caption{Neutron reflectivity spectra taken at 4.3 K in zero field
   and in 250 G (below the saturation value) for sample I $(a)$,
   and in zero and 185 G (well above the saturation value) for sample III 
   $(b)$.
   For sample I the applied intermediary field strength 250 G reduces the AFM
   peak intensity only by half, emerging FM peak is also visible.
   Data taken at 35 and 30 K (above $T_{\rm C}$) show
   the nonmagnetic contributions to the scattering.
   The reflectivity curves measured  in magnetic fields are shifted one
   order of  magnitude up for greater clarity.}
  \label{ng1_1}
  \end{figure}

The conclusion from the diffraction experiments is  further corroborated
by the results of the reflectometry studies. A reflectometer operating
on the NCNR NG-1 cold neutron guide was employed to get
the neutron reflectivity spectra.  Due to the longer neutron wavelength
($\lambda=4.75$ \AA), the $Q-$resolution in the reflectivity measurements was
considerably better than in the diffraction experiments.
Fig.~\ref{ng1_1} shows  the reflectivity profiles obtained for samples
I and III at 4.3~K.  In zero applied magnetic field,
pronounced peaks at the $Q_z$ position corresponding to twice
the chemical SL periodicity were observed  in both samples.
These maxima are again clear indication of the AFM alignement
of the magnetizations in successive EuS layers.
To confirm the magnetic origin of these peaks,   reflectivity spectra
were also taken with an in-plane magnetic field.
Application of a {\em sufficiently strong}, external magnetic field results
in full parallel alignment of the ferromagnetic \chem{EuS} layers; thus the AFM
peak disappears, while the intensity of the FM peak at the structural
position increases (Fig.~\ref{ng1_1}).  Details of the
AFM peak decay with applied magnetic field are discussed below.

\begin{figure}
  \twofigures[scale=0.7]{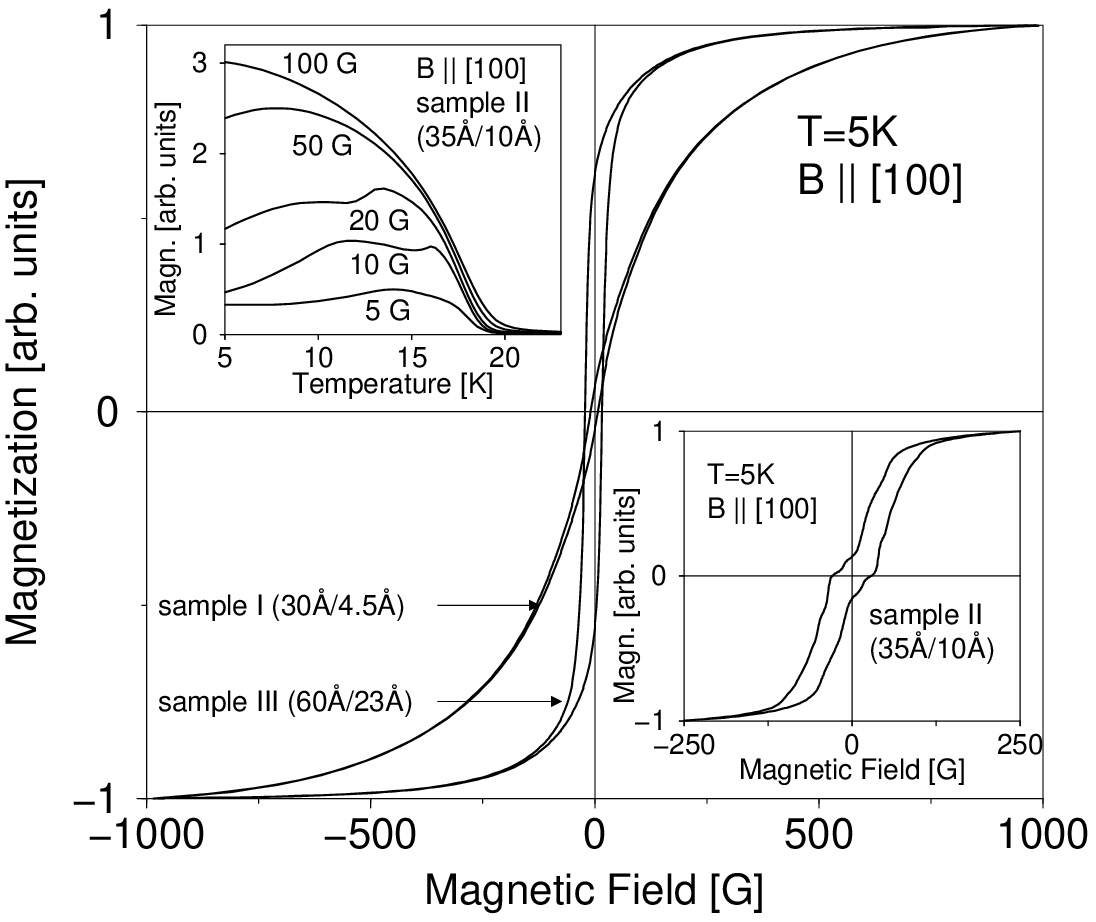}{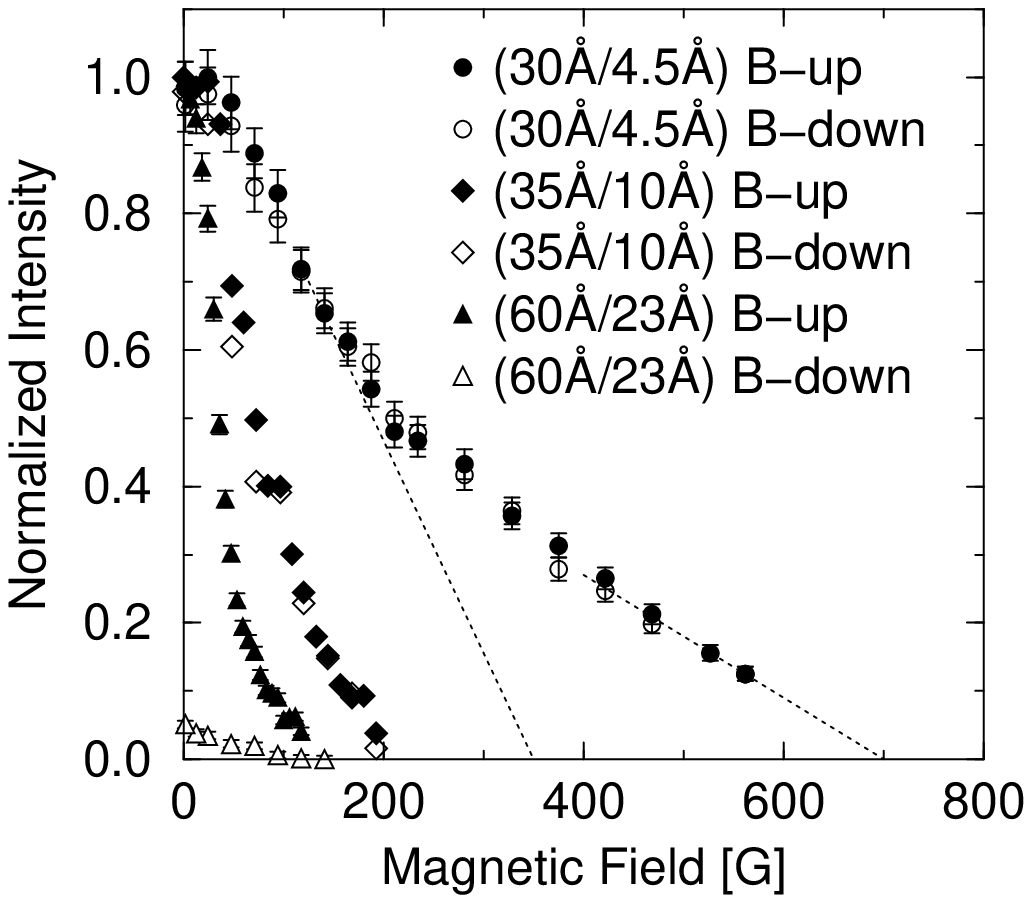}
  \caption {Field dependence of the magnetization of sample I and III
    along the crystallographic $[100]$ direction at 5 K is studied using 
    a cryogenic superconducting quantum interference device 
    (SQUID) magnetometer.  The lower-right 
    and upper-left  insets show  the magnetization  behaviour versus field  
    and temperature respectively of the sample II.}
  \label{mag_curves}
  \caption {Magnetic field dependence of the AFM reflectivity peak intensity
   for samples I, II, and III. Closed and open symbols represent the 
   data taken with increasing and decreasing fields respectively. The two
   slopes visible for the sample I, indicated by the dotted lines, 
   show the behavior of two regions in the specimen with the 1 and 2 
   monolayer thick \chem{PbS} spacer (see text).}
  \label{ng1_2}
  \end{figure}

Fig.~\ref{mag_curves} shows the  field  dependence of the
magnetization  taken by a SQUID magnetometer with the in-plane
field applied along the crystallographic $[100]$  direction at 5K.
The magnetic hysteresis loop for sample I shows a considerably harder
magnetic behaviour than for sample III,
indicating the interlayer coupling is stronger  with
the thinner \chem{PbS} spacer.  The lower remanence of Sample I
 --- 7$\%$ compared to 62$\%$ of Sample III ---
 clearly demonstrates that  the coupling is {\em antiferromagnetic}.
For sample II with intermediate spacer thicknesses,
a step-like behavior is  observed in the magnetization loop,
which is shown in the lower-right inset of Fig. \ref{mag_curves}.
The step-like behavior is a manifestation of the abrupt transition from
one stable magnetic arrangement of the EuS layers to another.
The upper-left inset shows an anomalous  temperature dependence of the same
sample ($d_{\scriptscriptstyle{\rm PbS}} = 10$\AA).  The decrease 
of the net  magnetization
with decreasing temperatures, observed for fields below 100~G, indicates
that AFM coupling becomes stronger at low temperatures.
For fields higher than 100~G, the temperature behavior of the magnetization
starts resembling that typically seen in ferromagnetic systems.  This is due to
the coupling being too weak to overcome the applied field strength for all 
temperatures down to 5K. 
The measurements taken along the $[110]$ crystallographic direction
show essentially the same field and temperature dependence for 
all the samples used in this study.

The results of  the field dependence studies of the AFM peak
in the neutron reflectivity spectrum are shown in Fig.~\ref{ng1_2} for 
all three samples. The measurement procedure starts with the samples in zero
magnetic field.
The field is gradually increased until the AFM peak disappears, and then is
decreased back to zero.
All samples show initially a strong AFM peak  before the application of the
magnetic field. To erase entirely
the AFM peak, a considerably larger fields are
needed for sample I ($\sim$ 700~G) and sample II ($\sim$ 200~G) 
than for sample III ($\sim$ 75~G). 
The AFM peak for sample I and sample II is {\em recoverable}.
The removal of the external field leads to the restoration of
the AFM coupled state.
The AFM peak of sample III is {\em not} recoverable.
The sample remains in the {\em ferromagnetic} configuration
after the removal of the external field. The irreversible destructive
effect of the magnetic field on the AFM reflectivity peak is consistent
with the fact that there are no traces of AFM coupling in the magnetic loop
measurements for sample III. This suggests that the AFM coupling strength in
sample III, with the thickest spacer, is much weaker than in the sample 
I or II. It is also weaker than the anisotropy (domain pinning) fields in 
sample III that lock the FM spin configuration after removal of the external 
magnetic field. Similar behavior of the AFM peak in external magnetic field has
been previously observed in Fe/Nb multilayers \cite{rehm}.

As noted, AFM coupling in all-semiconductor SL structures has not  been
predicted by previous theoretical models. In the
case of layers exhibiting a domain structure with small enough average domain
size (1$\mu$m or less), the dipolar coupling, proposed
in \cite{borchers} for metallic structures, may become sizable.
A characteristic feature of this
coupling is the relatively weak dependence of its strength on the spacer
thickness.
The mean domain size can be evaluated from the width of transverse
scans (wavevector transfer component parallel to the layers surface)
through the magnetic superlattice Bragg peaks.
In these scans performed for the sample I, a narrow,
specular peak is observed, its width and shape being almost identical
as that seen in the corresponding X-ray transverse scan. Thus the existing
broadening beyond
the instrumental resolution may be attributed to interface roughness
and other structural imperfections and not to the magnetic domain scattering.
Even if we attributed the total FWHM of this peak  to domain scattering, 
it would set a
lower limit on domain size of $\sim 3 \mu$m. The dipolar coupling strength
resulting from domains of this size with a 30 \AA{} EuS layer thickness is
lower than 80 G (see formula (9) in \cite{borchers}). Hence, the mechanism
might only contribute for spacer thicknesses larger than in the presented
samples. Traces of the interlayer coupling, slowly decreasing with the spacer
thickness, indeed observed in the neutronographic spectra for SLs up to
$d_{\scriptscriptstyle{\rm PbS}} =  90$ \AA{}, can be perhaps attributed to
the dipolar mechanism.
However, the dipolar forces cannot explain the behavior for
$d_{\scriptscriptstyle{\rm PbS}} < 25$ \AA{}, where the data show a rapid
increase of the coupling strength with decreasing PbS thickness ($\sim$ 700 G
for sample I), clearly suggesting that another stronger AFM interaction
takes over.

In search for a mechanism that could  explain that stronger interlayer
interaction, we analyzed the
sensitivity of the total energy of valence electrons to the magnetic
structure of a perfect \chem{(EuS)_m/(PbS)_n(001)} SL, where $m$ 
and $n$ are the
numbers of \chem{EuS} and \chem{PbS} monolayers respectively (the monolayer 
thickness of both
constituents is $\approx 3$~\AA{}). Specifically, we
calculated  the difference between the total electron energies of SLs with
parallel and antiparallel spin directions in consecutive \chem{EuS} layers. 
This
difference is a measure of the magnetic interlayer coupling. We used  an
empirical tight-binding method, which, although being a one-electron method,
should still describe adequately  the small spin-dependent changes in the
total energy.
The values of most tight-binding
parameters are chosen so as to reproduce the essential features of the
band structures of bulk \chem{EuS} \cite {Cho} and \chem{PbS} \cite {Wei},
the remaining being estimated by Harrison scaling.  Our calculations indicate
that the AFM spin alignment in successive \chem{EuS} layers
leads to a  lower energy of the SL system, in accord with
experimental observations.  The  coupling energy results primarily
from the \chem{Eu} ions at the magnetic/non-magnetic interfaces. The calculated
coupling strength  decreases with the thickness of the
non-magnetic spacer roughly like $2^{-n}$.  This is in a very good
agreement with the data presented in Fig.~\ref{ng1_2}, which allow to 
estimate the strength of the coupling, from the value of magnetic
field erasing the AFM neutron reflectivity peak (saturation field). 
A rough estimate of the  magnitude of coupling strength can be made 
using the formula $J_{1} = M_{s}Bt/4$ 
where $M_{s}$ is the saturation magnetization, and $t$ the thickness of
an EuS layer \cite{rehm}. 
As can be seen from the Fig.~\ref{ng1_2} these fields 
are $\sim$700~G, $\sim$350~G and $\sim$200~G  for
n=1, 2 and 3 respectively.  The first two values were deduced from the two
slopes for the sample I visible in Fig.~\ref{ng1_2}. 
As 4.5~\AA{} spacer is obout 1.5 
monolayers thick, the sample I may be considered as a mixture of two 
superlattices with $n=1$ and $2$.
Theoretically calculated values
of $J_{1}$ for the {\em perfect} SL with n=1, 2 and 3 -- 0.77, 0.33,
and 0.18 mJ/m$^2$, respectively -- although follow the same PbS spacer 
thickness dependence are about an 
order of magnitude higher than the experimental values estimated from 
the saturation fields using the above formula (0.063, 0.031, and 
0.019 mJ/m$^2$ respectively). One of the possible reasons for the latter 
discrepancy may be interfacial roughness and interdiffusion.
Substancial reduction of the coupling strength in metallic multilayers due
to the alloying effects in the interfacial regions have previously 
been reported \cite{bruno2}. The full, detailed description of the above 
theoretical calculations can be found in \cite{blin2}.

In summary, we have presented the first conclusive experimental evidence,
based on
neutron diffraction, neutron reflectivity, and magnetization measurements,
for the existence of antiferromagnetic interlayer exchange coupling in a
purely semiconductor, ferromagnetic/nonmagnetic epitaxial
system with negligible carrier concentration. We have also shown that
the experimentally observed sign, strength and range of the interaction may be
well accounted for by band structure effects sensitive to the magnetic
configuration of the multilayer.
\acknowledgments
Acknowledgments: Work supported by projects: NATO PST.CLG 975228,
NSF DMR-9972586, KBN 2 P03B 007 16, and KBN  2 P03B 154 18.

\end{document}